\documentclass[preprint,prd,showpacs,amsmath,amssymb,amsthm,nofootinbib]{revtex4}
\usepackage[mathscr]{euscript}
\usepackage{bm}
\usepackage{graphicx}
\usepackage{subfigure}
\usepackage{multirow}
\usepackage[colorlinks=true,linkcolor=red]{hyperref}
\newcommand{\bea}{\begin{eqnarray}}
\newcommand{\eea}{\end{eqnarray}}
\newcommand{\beq}{\begin{equation}}
\newcommand{\eeq}{\end{equation}}

\def\/{\over}

\begin{document}

\title{Emergent scenario in mimetic gravity}
\author{Qihong Huang$^1$\footnote{Corresponding author: huangqihongzynu@163.com}, Bing Xu$^2$\footnote{Corresponding author: xub@ahstu.edu.cn}, He Huang$^3$, Feiquan Tu$^1$, Ruanjing Zhang$^4$}
\affiliation{
$^1$School of Physics and Electronic Science, Zunyi Normal University, Zunyi 563006, People's Republic of China\\
$^2$School of Electrical and Electronic Engineering, Anhui Science and Technology University, Bengbu 233030, People's Republic of China\\
$^3$College of Physics and Electronic Engineering, Nanning Normal university, Nanning 530001, People's Republic of China\\
$^4$College of Science, Henan University of Technology, Zhengzhou 450001, People's Republic of China
}

\begin{abstract}
The emergent scenario provides a possible way to avoid the big bang singularity by assuming that the universe originates from an Einstein static state. Therefore, an Einstein static universe stable under perturbations is crucial to a successful implementation of the emergent mechanism. In this paper, we analyze the stability of the Einstein static universe against the scalar perturbations in the mimetic theory and find that stable Einstein static solutions exist under certain conditions in this theory. In the original mimetic gravity, the Einstein static universe is unstable. Then, we find that the universe can naturally exit from the initial static state, evolve into an inflationary era and then exit from the inflationary era. Thus, the emergent scenario can be used to resolve the big bang singularity in the mimetic theory.
\end{abstract}

\maketitle

\section{Introduction}

Inflation~\cite{Guth1981} is a short-lived and prompt accelerated expansion era in the early universe, which can solve most of problems in the standard cosmology. However, it still suffers from the big bang singularity problem. In order to resolve this initial singularity of the universe, some scenarios have been suggested to construct non-singular or past eternal cosmological models, such as the pre-big bang~\cite{Gasperini2003, Lidsey2000}, the cyclic scenario~\cite{Khoury2001, Steinhardt2002, Khoury2004} and the emergent universe ~\cite{Ellis2004}. The emergent scenario is introduced first by Ellis $et$ $al.$~\cite{Ellis2004, Ellis2004a} in the context of general relativity, which can circumvent the big bang singularity by assuming that the universe originates from a past eternal Einstein static (ES) universe. So, a stable ES solution against perturbations is crucial to successful implementation of emergent scenario. Unfortunately, the ES solutions in general relativity~\cite{Barrow2003, Eddington1930, Gibbons1987, Gibbons1988} is unstable against scalar perturbations, and as a result, the emergent scenario can not be successfully realized as expected in general relativity. Therefore, lots of efforts have been made to go beyond general relativity and the stable ES state solutions have been found in some theories of modified gravity~\cite{Carneiro2009, Huang2014, Campo2007, Campo2009, Wu2011, Li2013, Mulryne2005, Lidsey2004, Parisi2007, Canonico2010, Wu2009, Bag2014, HuangQ2015, Bohmer2004, Atazadeh2014, Zhang2016, Zhang2010, Zhang2012, Lidsey2006, Gruppuso2004, Gergely2002, Atazadeh2014a, Zhang2014, Heydarzade2016, Heydarzade2015, HuangH2015, Bohmer2009, Wu2010, Bohmer2010, Khodadi2016, Bohmer2015, Atazadeh2015, Atazadeh2017, Bohmer2013, Tawfik2016, Khodadi2015, Odrzywolek2009, Clifton2005, Vilenkin2013, Aguirre2013, Mithani2012, Cai2014, Cai2012, Mousavi2017, Miao2016, Seahra2009, Li2017, Shabani2017, Huang2018, HuangQ2018, Shabani2019, Sharif2019, Li2019}.

Recently, Chamseddine and Mukhanov~\cite{Chamseddine2013} proposed a new modified gravitational theory, dubbed as mimetic gravity, which is an extension of general relativity. In this theory, the physical metric is defined in terms of an auxiliary metric $\tilde{g}_{\mu\nu}$ and  a scalar field $\phi$. The resulting field equations differ from Einstein's equations, due to the appearance of an extra mode of the gravitational field which can be treated as a source of dark matter. After a Lagrange multiplier was introduced, the equivalent formulation of the mimetic theory was obtained without the auxiliary metric $\tilde{g}_{\mu\nu}$~\cite{Golovnev2014, Chamseddine2014}. Although this theory was first proposed as a model of dark matter, it was subsequently realized that it can serve as a model of dark energy after introducing the potential of the scalar field. It also provides an inflationary mechanism and a bouncing universe~\cite{Chamseddine2014}. This stimulated some interests in the literature, and its relation with disformal transformations was given in ~\cite{Deruelle2014, Domenech2015, Arroja2015} and the Hamiltonian analyses were discussed in ~\cite{Malaeb2015, Langlois2016, Takahashi2017, Ganz2019}. Other works on mimetic theory, phenomenology, and its observational viability were examined in~\cite{Arroja2016, Chaichian2014, Leon2015, Ramazanov2015, Matsumoto2015, Mirzagholi2015, Ijjas2016, Nojirie2016, Momeni2016, Saadi2016, Zheng2017, Firouzjahi2017, Hirano2017, Sadeghnezhad2017, Langlois2019, Casalino2019, Chamseddine2019, Nojiri2014, Myrzakulov2015, Chamseddine2016, Vagnozzi2017, Babichev2017, Dutta2018, Astashenok2016, Myrzakulov2016, Arroja2018, Casalino2018, Ganz2019a}. In addition, a review of mimetic gravity was given in~\cite{Sebastiani2017}. However, when the model was applied to early universe, it failed to explain the origin of the large scale structure from quantum fluctuations since the speed of sound vanishes during the perturbations. As a result, its fluctuations cannot be quantized in a usual way. To resolve this problem, mimetic theory with higher derivative terms~\cite{Chamseddine2014, Zheng2017}, mimetic Horndeski gravity~\cite{Cognola2016}, nonminimal extension of mimetic matter~\cite{Casalino2018, Hosseinkhan2018}, mimetic scalar-tensor theories~\cite{Ganz2019}, mimetic f(T) gravity~\cite{Guo2020}, and so on, are later presented.

In this paper, we discuss whether the emergent mechanism can be used to avoid the big bang singularity in mimetic theory in which mimetic scalar field is a non-minimal kinetic term coupled to the gravity. By analyzing the stability of ES solutions against scalar perturbations, we obtain the stability conditions of ES solutions. Then, we find that the universe can naturally exit from initially stable static state, evolve into an inflationary era and then exit from inflation. The paper is organized as follows. In Section II, we give the field equations of mimetic theory and discuss ES solutions. In Section III, the stability of ES solutions against scalar perturbations is analyzed. In Section IV, we discuss how the universe exits from the ES state, evolves to an inflationary era and exits from inflation. Finally, our main conclusions are presented in Section V. Throughout this paper, unless specified, we adopt the metric signature ($-, +, +, +$). Latin indices run from 0 to 3 and the Einstein convention is assumed for repeated indices. In this paper, we adopt $8\pi G=c=1$.

\section{The field equations and Einstein static solutions}

We start with the following action for a cosmological model in which mimetic scalar field is a non-minimal kinetic term coupled to the gravity~\cite{Cognola2016, Casalino2018}
\bea\label{action}
S=\int d^{4}x \sqrt{-g}\Big[\frac{1}{2}f(X)R+\lambda(g^{\mu\nu}\partial_{\mu}\phi \partial_{\nu}\phi+1)-V(\phi)+\frac{1}{2}\gamma(\square\phi)^{2}\Big]+S_{m},
\eea
where $g_{\mu\nu}$ is the metric tensor with $g$ being its determinant, $R$ denotes the Ricci curvature scalar, $X=\frac{1}{2}g^{\mu\nu}\partial_{\mu}\phi\partial_{\nu}\phi$, $f(X)=1-2\xi X$ shows the non-minimal coupling between the kinetic term of the mimetic scalar field and the Ricci curvature scalar, $\xi$ is the coupling constant, $\lambda$ stands for the Lagrange multiplier, $\square=g^{\mu\nu}\nabla_{\mu}\nabla_{\nu}$,  $S_{m}$ represents the action of a perfect fluid, $\gamma$ is a constant. $\xi=0$ and $\gamma=0$ corresponds to the case of the original mimetic theory. $V(\phi)$ is the mimetic potential, we assume it taking the form $V=V_{0}+e^{\phi-\alpha}+\beta[\tanh(\sigma-\phi)-1]$ which is a constant $V_{0}$ at $\phi\rightarrow -\infty$ and that is how we can extract the the Einstein static solution.

\subsection{Field equations}

Varying the action~(\ref{action}) with respect to the metric tensor $g_{\mu\nu}$, the scalar field $\phi$ and the Lagrange multiplier $\lambda$, respectively, one can obtain:
\bea\label{Einstein}
&&G_{\mu\nu}(1-\xi g^{\alpha\beta}\nabla_{\alpha}\phi\nabla_{\beta}\phi)-\xi R\nabla_{\mu}\phi\nabla_{\nu}\phi=-\Big[V+\gamma\nabla_{\lambda}\phi \nabla^{\lambda}(\square\phi)+\frac{1}{2}\gamma(\square\phi)^{2}\Big]g_{\mu\nu}\nonumber\\
&&\qquad\quad-2\lambda\partial_{\mu}\phi \partial_{\nu}\phi+\gamma\big[\nabla_{\nu}\phi \nabla_{\mu}(\square\phi)+\nabla_{\mu}\phi \nabla_{\nu}(\square\phi)\big]+T_{\mu\nu},
\eea
\bea\label{scalar}
\xi\nabla^{\mu}(R\nabla_{\mu}\phi)-2\nabla^{\mu}(\lambda\nabla_{\mu}\phi)
 +\square(\gamma\square\phi)=0,
\eea
and
\bea\label{Lagrange}
g^{\mu\nu}\partial_{\mu}\phi\partial_{\nu}\phi+1=0,
\eea
where $T_{\mu\nu}$ represents the energy-momentum tensor of a perfect fluid.

To find an ES solution, we consider a homogeneous and isotropic closed universe described by the Friedmann-Lema$\hat{\imath}$tre-Robertson-Walker metric
\bea
ds^{2}=-dt^{2}+ a(t)^2 \gamma_{ij} dx^{i} dx^{j},
\eea
where $a(t)$ denotes the scale factor, and $\gamma_{ij}$ represents the metric on the three-sphere
\bea
\gamma_{ij} dx^{i} dx^{j}=\frac{dr^{2}}{1-r^{2}}+r^{2}(d\theta^2+\sin^{2}\theta d\phi^2) .
\eea
From Eq.~(\ref{Lagrange}), we obtain
\bea\label{L}
\dot{\phi}^{2}-1=0.
\eea
The ($00$) and ($ij$) components of Eq.~(\ref{Einstein}) give
\bea\label{00}
3\Big(H^{2}+\frac{1}{a^{2}}\Big)(1+\xi \dot{\phi}^{2})-6\xi\Big(2H^{2}+\dot{H}+\frac{1}{a^{2}}\Big)\dot{\phi}^{2}=V+\gamma\dot{\varphi}\dot{\phi}+\frac{1}{2}\gamma\varphi^{2}-2\lambda\dot{\phi}^{2}+\rho,
\eea
\bea\label{ij}
\Big(3H^{2}+2\dot{H}+\frac{1}{a^{2}}\Big)(1+\xi \dot{\phi}^{2})=V-\gamma\dot{\varphi}\dot{\phi}+\frac{1}{2}\gamma\varphi^{2}-p,
\eea
with
\bea
\varphi=\square\phi=-\ddot{\phi}-3H\dot{\phi},\nonumber
\eea
where $\rho$ and $p$ are the energy density and pressure of a perfect fluid, respectively, which satisfy $p=w\rho$ with $w$ being the  equation of state, $H=\frac{\dot{a}}{a}$ and a dot denotes the derivative with respect to $t$, where $t$ is cosmic time. From Eq.~(\ref{scalar}), we have the following dynamical equation of scalar field
\bea\label{s}
-6\xi\Big(4H\dot{H}+\ddot{H}-2H\frac{1}{a^{2}}\Big)\dot{\phi}+6\xi\Big(2H^{2}+\dot{H}+\frac{1}{a^{2}}\Big)\varphi+2\dot{\lambda}\dot{\phi}-2\lambda\varphi-\gamma(\ddot{\varphi}+3H \dot{\varphi})=0.
\eea

\subsection{Einstein static solutions}

For the ES solutions, the static conditions $\dot{a}=\ddot{a}=0$ imply $a=a_{0}=constant$ and $\dot{H}=H=0$. And the mimetic constraint $\dot{\phi}^{2}-1=0$ gives $\ddot{\phi}=0$ and $\varphi=0$. Using these conditions, Eqs.~(\ref{L}), ~(\ref{00}), ~(\ref{ij}) and ~(\ref{s}) can be reduced as
\beq\label{ss1}
\dot{\phi}^{2}_{0}=1,
\eeq
\beq\label{ss2}
3(1-\xi)\frac{1}{a^{2}_{0}}=V_{0}-2\lambda_{0}+\rho_{0},
\eeq
\beq\label{ss03}
(1+\xi)\frac{1}{a^{2}_{0}}=V_{0}-p_{0},
\eeq
\beq
2\dot{\lambda}_{0} =0,
\eeq
where the subscript $0$ stands for the value at ES state. Obviously, $V_{0}$, $\lambda_{0}$ and $\rho_{0}$ should be constants in ES state.

Combining Eqs.~(\ref{ss2}) and (\ref{ss03}), we can obtain
\bea
&&\frac{1}{a^{2}_{0}}=\frac{(1+\omega)V_{0}-2\omega \lambda_{0}}{1+3\omega+(1-3\omega)\xi},\\
&&\rho_{0}=\frac{2[\lambda_{0}+V_{0}+(\lambda_{0}-2V_{0})\xi]}{1+3\omega+(1-3\omega)\xi}.
\eea
Since $a_{0}$ and $\rho_{0}$ are positive values, the existence conditions of ES solutions are $a^{2}_{0}>0$ and $\rho_{0}>0$. So, the existence conditions of ES solutions are
\beq
\omega\leq 0, \quad -1<\xi<\frac{3\omega+1}{3\omega-1}, \quad \lambda_{0}<\frac{(2\xi-1)V_{0}}{1+\xi};\\ \nonumber
\eeq
\beq
\omega\leq 0, \quad \xi>\frac{3\omega+1}{3\omega-1}, \quad \lambda_{0}>\frac{(2\xi-1)V_{0}}{1+\xi};\\ \nonumber
\eeq
\beq
0<\omega<\frac{1}{3}, \quad \xi<\frac{3\omega+1}{3\omega-1}, \quad \lambda_{0}>\frac{(1+\omega)V_{0}}{2\omega};\\ \nonumber
\eeq
\beq
0<\omega<\frac{1}{3}, \quad \frac{3\omega+1}{3\omega-1}<\xi\leq -1, \quad \lambda_{0}<\frac{(1+\omega)V_{0}}{2\omega};\\ \nonumber
\eeq
\beq
0<\omega<\frac{1}{3}, \quad \xi> -1, \quad \frac{(2\xi-1)V_{0}}{1+\xi}<\lambda_{0}<\frac{(1+\omega)V_{0}}{2\omega};\\ \nonumber
\eeq
\beq
\omega=\frac{1}{3}, \quad \xi\leq -1, \quad \lambda_{0}< 2V_{0};\\ \nonumber
\eeq
\beq
\omega=\frac{1}{3}, \quad \xi> -1, \quad \frac{(2\xi-1)V_{0}}{1+\xi}<\lambda_{0}< 2V_{0};\\ \nonumber
\eeq
\beq
\omega>\frac{1}{3}, \quad \xi\leq -1, \quad \lambda_{0}<\frac{(1+\omega)V_{0}}{2\omega};\\ \nonumber
\eeq
\beq
\omega>\frac{1}{3}, \quad -1<\xi<\frac{3\omega+1}{3\omega-1}, \quad \frac{(2\xi-1)V_{0}}{1+\xi}<\lambda_{0}<\frac{(1+\omega)V_{0}}{2\omega};\\ \nonumber
\eeq
\beq
\omega>\frac{1}{3}, \quad \xi>\frac{3\omega+1}{3\omega-1}, \quad \frac{(1+\omega)V_{0}}{2\omega}<\lambda_{0}<\frac{(2\xi-1)V_{0}}{1+\xi}.
\eeq
In the following, we will discuss the stability of ES solutions under the scalar perturbations.

\section{Scalar perturbations}

In this section, we will analyze the effect of scalar perturbations on the stability of ES solutions. The perturbed metric takes the form ~\cite{Bardeen1980}
\bea\label{ds1}
ds^{2}=-(1+2\Psi)dt^{2}+ a(t)^{2}(1+2\Phi) \gamma_{ij} dx^{i} dx^{j},
\eea
where the Newtonian gauge has been used, $\Psi$ and $\Phi$ represent the Bardeen potential and the perturbation to the spatial curvature, respectively. For the perfect fluid, the perturbation of energy-momentum tensor is
\bea
\delta T^{\mu}_{\nu}=(1+c^{2}_{s}) u^{\mu}_{\nu} \delta\rho+(1+w)\rho(\delta u^{\mu}u_{\nu}+u^{\mu}\delta u_{\nu})+c_{s}^{2} \delta^{\mu}_{\nu} \delta\rho,\nonumber
\eea
where $u^{\mu}$ is the four-velocity. Neglecting the perturbations are higher than the first order, we obtain
\bea
&&u^{\mu}=\big(1-\Psi,\frac{v^{i}}{a}\big),\nonumber\\
&&u_{\mu}=\big(-1-\Psi,a v_{i}\big),\nonumber\\
&&u_{\mu}u^{\mu}=-1,\nonumber
\eea
where $v$ is a velocity potential related to $v^{i}$ via $v^{i}=\nabla^{i}v$. The relation between density and pressure perturbations is
\beq
\delta p=c^{2}_{s} \rho_{0}\delta+3\mathcal{H}(1+\omega)(c^{2}_{s}-c^{2}_{a})\rho_{0}\frac{\theta}{k^{2}},\nonumber
\eeq
where $c^{2}_{a}$ is the adiabatic sound speed
\beq
c^{2}_{a}=\omega-\frac{\omega'}{3\mathcal{H}(1+\omega)},\nonumber
\eeq
and $\theta$ is the velocity divergence. By assuming that $\omega$ is constant and $c^{2}_{s}=c^{2}_{a}$, we obtain
\bea\label{rhop}
\delta p=c^{2}_{s} \rho_{0}\delta.
\eea
Here $\delta=\delta\rho/\rho_{0}$ and the sound speed $c^{2}_{s}=\omega$.

Combining the perturbed metric~(\ref{ds1}) and the field equations given in Eqs.~(\ref{Einstein}),~(\ref{scalar}) and ~(\ref{Lagrange}), we have the following perturbation equations
\bea\label{p1}
&&\frac{2}{a^{2}_{0}}\nabla^{2}\Phi+\frac{6}{a^{2}_{0}}\Phi+\Big[\frac{6}{a^{2}_{0}}\delta\dot{\phi}-\frac{2}{a^{2}_{0}}\nabla^{2}(\Phi+\Psi)+6\ddot{\Phi}-\frac{6}{a^{2}_{0}}(\Phi+\Psi)\Big]\xi\\ \nonumber
&&\qquad\quad=4\lambda_{0}(\delta\dot{\phi}-\Psi)-\gamma\delta\dot{\varphi}+2\delta\lambda-\delta\rho,
\eea
\beq
2\dot{\Phi}(1+\xi)=2\big(\lambda_{0}-3\xi\frac{1}{a^{2}_{0}}\big)\delta\phi-\gamma\delta\varphi+q,
\eeq
\beq\label{p2}
\Psi=-\Phi,
\eeq
\beq\label{p3}
\Big[6\Big(-\ddot{\Phi}+\frac{1}{a^{2}_{0}}\Phi\Big)+\frac{2}{a^{2}_{0}}\nabla^{2}(\Psi+\Phi)\Big](1+\xi)+\frac{6}{a^{2}_{0}}(\Psi-\delta\dot{\phi})\xi=3\gamma\delta\dot{\varphi}+3\delta p,
\eeq
\beq\label{p4}
\xi(-\delta\dot{R}+R \delta\varphi)+2\delta\dot{\lambda}-2\lambda_{0}\delta\varphi-\gamma\delta\ddot{\varphi}+\gamma\frac{1}{a^{2}_{0}}\nabla^{2}\delta\varphi=0,
\eeq
\beq\label{p5}
\Psi-\delta\dot{\phi}=0,
\eeq
where
\bea\label{p6}
&&\delta\varphi=\delta(\square\phi)=-\delta\ddot{\phi}+\frac{1}{a^{2}_{0}}\nabla^{2}\delta\phi+\dot{\Psi}-3\dot{\Phi},\nonumber\\
&&\delta R=6\ddot{\Phi}-\frac{2}{a^{2}_{0}}\nabla^{2}\Psi-\frac{4}{a^{2}_{0}}\nabla^{2}\Phi-\frac{12}{a^{2}_{0}}\Phi.
\eea
Here, the perturbation of the scalar field $\phi\rightarrow \phi_{0}+\delta \phi$ and Lagrange multiplier $\lambda\rightarrow \lambda_{0}+\delta \lambda$ are considered, and $q=a(\rho+p)v$ is the velocity perturbation.

From the perturbation of the equation of continuity, $\nabla_{\mu}T^{\mu}_{\nu}=0$, one obtains
\bea
&&\dot{\delta\rho}+\frac{1}{a^{2}_{0}}\nabla^{2}q+3(\rho+p)\dot{\Phi}=0,\nonumber\\
&&\dot{q}+c^{2}_{s}\delta\rho+(\rho+p)\Psi=0.\nonumber
\eea

For convenience, we can perform a harmonic decompositions for all perturbation terms
\bea\label{Y}
&&\Psi=\Psi_{nlm}(t)Y_{nlm}(\theta^{i}), \quad \Phi=\Phi_{nlm}(t)Y_{nlm}(\theta^{i}), \quad q=q_{nlm}(t)Y_{nlm}(\theta^{i}),\nonumber\\
&&\delta=\delta_{nlm}(t)Y_{nlm}(\theta^{i}), \quad \delta\phi=\delta\phi_{nlm}(t)Y_{nlm}(\theta^{i}), \quad \delta\lambda=\delta\lambda_{nlm}(t)Y_{nlm}(\theta^{i}),\nonumber
\eea
where summations over $n, m, l$ are implied. The quantum numbers $m$ and $l$ will be suppressed hereafter since they do not enter the differential equation for the scalar perturbations. The harmonic function $Y_{n}=Y_{nlm}(\theta^{i})$ satisfies~\cite{Harrison1967}
\bea\label{Yn}
\Delta Y_{n}=-k^{2}Y_{n}=\Bigg\{
\begin{array}{rrr}
-n(n+2)Y_{n}, \quad n=0,1,2,..., \quad K=+1\\
-n^{2}Y_{n}, \quad\quad\quad\quad n^{2}\geq 0, \quad\quad\quad\quad K=0\\
-(n^{2}+1)Y_{n}, \quad\quad n^{2}\geq 0, \quad\quad\quad K=-1
\end{array}
\eea
Here, $\Delta$ is the 3-dimensional spatial Laplacian operator, $k=0(n=0)$ corresponds to the case of homogeneous scalar perturbations, $K=+1, 0$ and $-1$ corresponds to a spatially closed, flat, and open universe, respectively. For $K=1$, the spectrum of the perturbation modes is discrete. While, it is continuous for $K=0$ or $-1$.

Then, using Eqs.~(\ref{p1}),~(\ref{p2}),~(\ref{p3}),~(\ref{p4}),~(\ref{p5}),~(\ref{p6}) and the static conditions, we obtain two perturbed equations
\beq\label{p11}
\delta\dddot{\Phi}_{n}+\alpha\dot{\Phi}_{n}-\beta\delta\phi_{n}=0,
\eeq
\beq\label{p12}
\delta\dot{\phi}_{n}+\Phi_{n}=0,
\eeq
with
\bea
&&\alpha=\frac{1}{2-3\gamma+2\xi}\Big[(\gamma+2w-3w\gamma+2w\xi)\frac{k^{2}}{a^{2}_{0}}-2(-6w\xi+\xi+3w+1)\frac{1}{a^{2}_{0}}
-6w\lambda_{0}\Big],\nonumber\\
&&\beta=\frac{1}{2-3\gamma+2\xi}\Big(2w\lambda_{0}+w\gamma\frac{k^{2}}{a^{2}_{0}}-6w\xi\frac{1}{a^{2}_{0}}\Big)\frac{k^{2}}{a^{2}_{0}}.
\eea
Since the theory contains higher derivative terms, one may wonder that the theory might be unstable due to the Ostrogradsky ghost. Usually the Ostrogradsky ghost arises when the higher derivative terms increase the number of degrees of freedom for this system under consideration. However, in the mimetic scenario, the propagation of an extra Ostrogradsky-like degree of freedom can be prevented since the mimetic constraint~(\ref{L}) removes one dynamical degree~\cite{Firouzjahi2017,Zheng2017}. As a result, the mimetic gravity does not suffer from the Ostrogradsky ghost.

After introducing two new variables $A=\dot{\Phi}$ and $B=\dot{A}$, Eqs.~(\ref{p11}) and ~(\ref{p12}) can be rewritten as
\bea
&&\dot{\Phi}_{n}-A_{n}=0,\nonumber\\
&&\dot{A}_{n}-B_{n}=0,\nonumber\\
&&\dot{B}_{n}+\alpha A_{n}-\beta \delta\phi_{n}=0,\nonumber\\
&&\delta\dot{\phi}_{n}+\Phi_{n}=0.
\eea
Then, the stability of ES solutions is determined by the eigenvalues of the coefficient matrix of this dynamical system, which is
\bea
\mu^{2}=-\frac{\alpha\pm\sqrt{\alpha^{2}-4\beta}}{2}.
\eea
For $\mu^{2}<0$, the ES solution is a center equilibrium point, and a small perturbation from the ES state will result in an oscillation around this static state rather than an exponential deviation from the static state. Thus, the ES solution is stable. Otherwise, it is unstable. That is to say, the stability under scalar perturbations is determined by $\mu^{2}<0$, which leads to
\bea\label{MN}
\alpha>0, \quad \alpha^{2}-4\beta>0, \quad 4\beta>0.
\eea
Since $\beta=0$ when $k^{2}=0$, the existence of stable ES solutions under homogeneous scalar perturbations requires
\bea\label{M}
\alpha>0.
\eea

\subsection{homogeneous perturbations}

In order to discuss the stability of ES solutions for scalar perturbations, we will analyze it under the existence conditions $a^{2}_{0}>0$ and $\rho_{0}>0$. In addition, the conditions $\frac{(1+\xi)(2-3\gamma+2\xi)}{\gamma}<0$ and $1+\xi<0$~\cite{Casalino2018}, which can be used to avoid the ghost and gradient instabilities, are also considered.

From Eq.~(\ref{M}), we can obtain that the stability conditions under homogeneous scalar perturbations which correspond to the case $n=0$ are
\beq\label{h1}
0<\omega<\frac{1}{3}, \quad \xi<\frac{3\omega+1}{3\omega-1}, \quad \frac{2}{3}(1+\xi)<\gamma<0, \quad \lambda_{0}>\frac{(1+\omega)V_{0}}{2\omega};
\eeq
\beq\label{h2}
0<\omega<\frac{1}{3}, \quad \frac{3\omega+1}{3\omega-1}<\xi<-1, \quad \frac{2}{3}(1+\xi)<\gamma<0, \quad \lambda_{0}<\frac{(1+\omega)V_{0}}{2\omega};
\eeq
\beq\label{h3}
\omega\geq\frac{1}{3}, \quad \xi<-1, \quad \frac{2}{3}(1+\xi)<\gamma<0, \quad \lambda_{0}<\frac{(1+\omega)V_{0}}{2\omega}.
\eeq

\subsection{inhomogeneous perturbations}

For inhomogeneous scalar perturbations, since the $n=1$ mode corresponds to a gauge degree of freedom related to a global rotation, which reflects the freedom to change the four-velocity of fundamental observers~\cite{Li2017}. And the physical modes have $n\geq 2$ which correspond to $k^{2}\geq 8$. Since a general analysis is too complicated, in the following, we will only examine inhomogeneous scalar perturbations under the conditions given by the homogeneous perturbations.

(i)For the condition Eq.~(\ref{h1}), the stability conditions ~(\ref{MN}) require
\bea
&&k^{2}\geq10, \quad V_{0}>0, \quad \frac{3}{2k^{2}-9}<\omega<\frac{1}{3}, \quad \xi<\frac{3\omega+1}{3\omega-1},\nonumber\\
&&\quad \frac{2}{3}(1+\xi)<\gamma<\frac{(1+\xi)(1+3\omega)}{\omega k^{2}}, \quad \lambda_{0}>\frac{(\gamma k^{2}-6\xi)(1+\omega)V_{0}}{2\omega\gamma k^{2}-2(1+\xi)(1+3\omega)}.
\eea
For inhomogeneous perturbations ($k^{2}\geq8$), the stable ES solutions are required to be stable for any $k$. As a result, the ES solutions are unstable in this case.

(ii)Combing the conditions Eq.~(\ref{h2}) and ~(\ref{MN}), one obtain
\bea
&&k^{2}\geq10, \quad V_{0}>0, \quad \frac{3}{2k^{2}-9}<\omega<\frac{1}{3}, \quad \frac{3\omega+1}{3\omega-1}<\xi<-1,\nonumber\\
&&\quad \frac{2}{3}(1+\xi)<\gamma<\frac{(1+\xi)(1+3\omega)}{\omega k^{2}}, \quad \lambda_{0}<\frac{(\gamma k^{2}-6\xi)(1+\omega)V_{0}}{2\omega\gamma k^{2}-2(1+\xi)(1+3\omega)},
\eea
which conflicts with $k^{2}\geq 8$. So, the ES solutions are also unstable in this case.

\begin{figure}[!htb]
                \centering
                \includegraphics[width=0.45\textwidth]{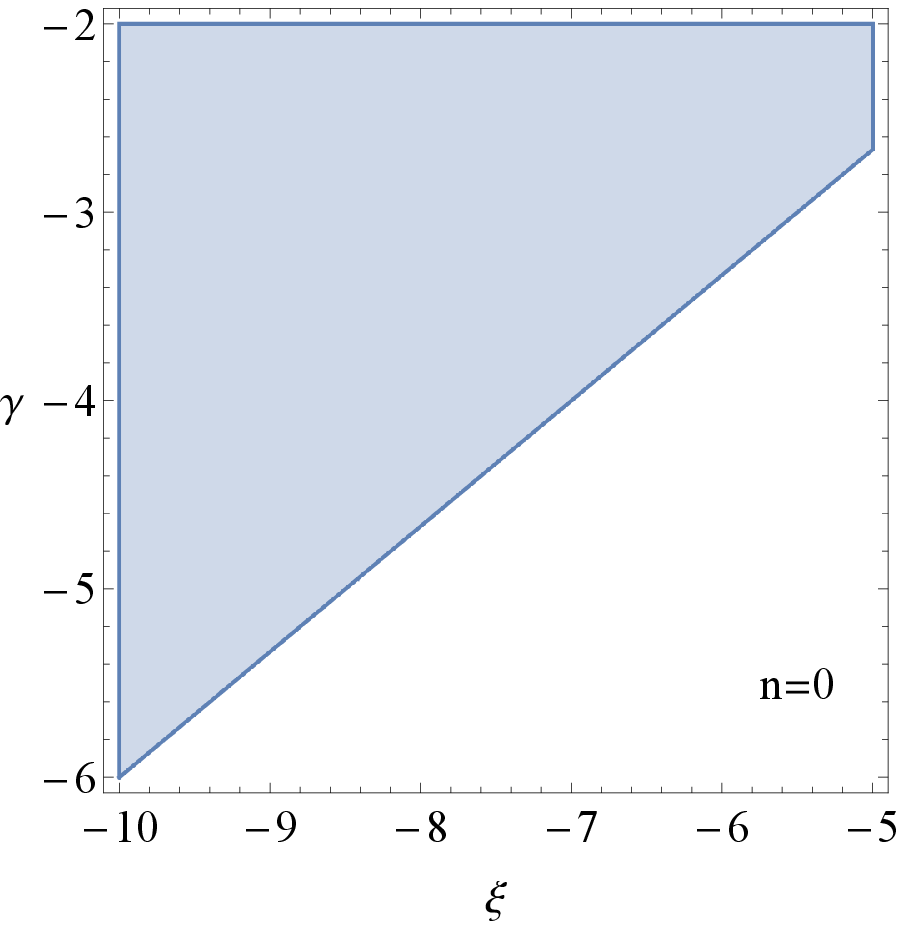}
                \includegraphics[width=0.45\textwidth]{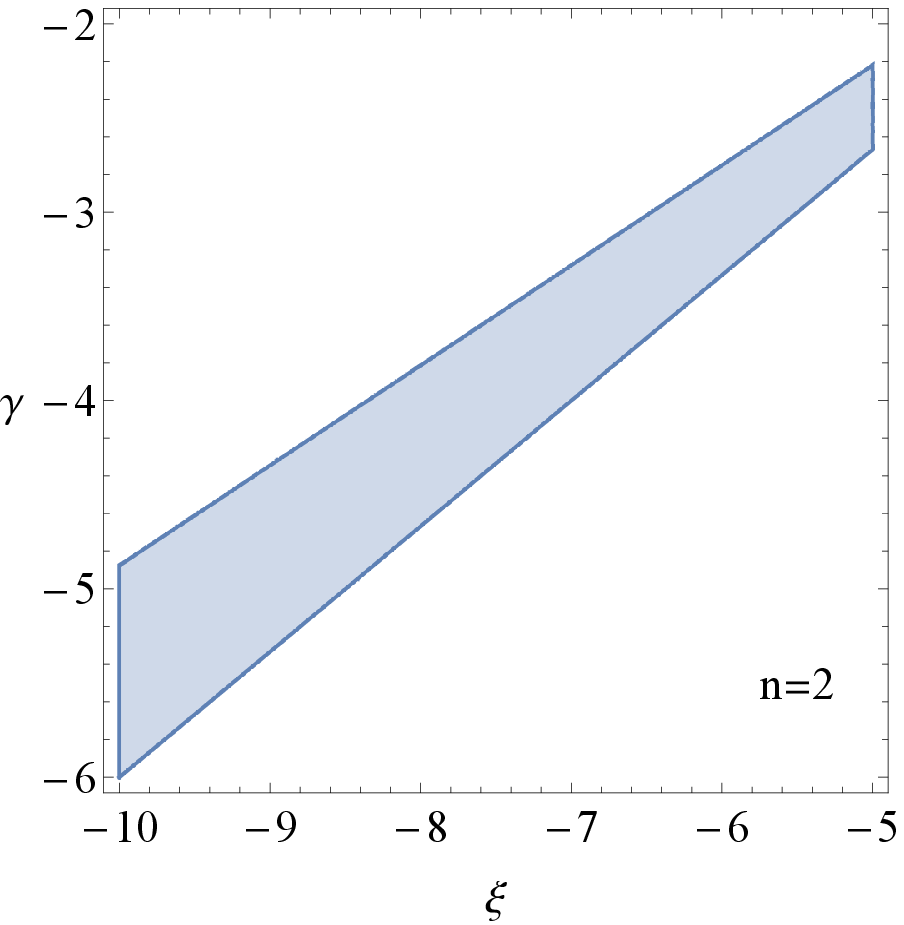}
                \includegraphics[width=0.45\textwidth]{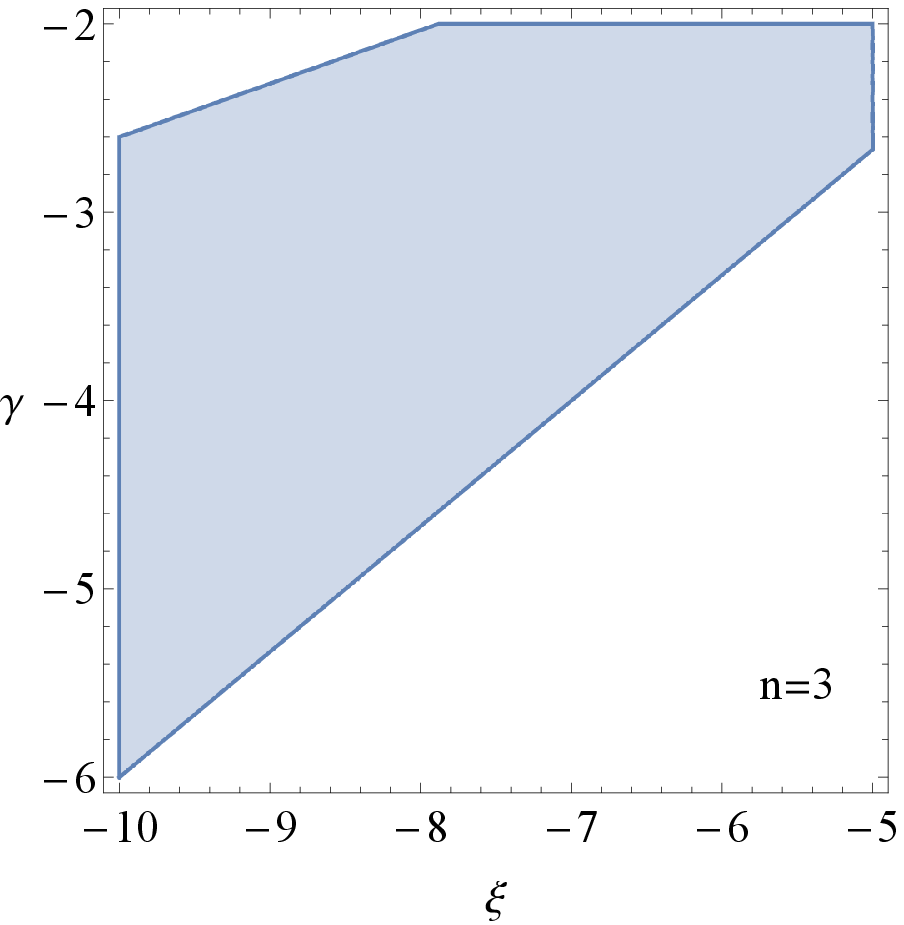}
                \includegraphics[width=0.45\textwidth]{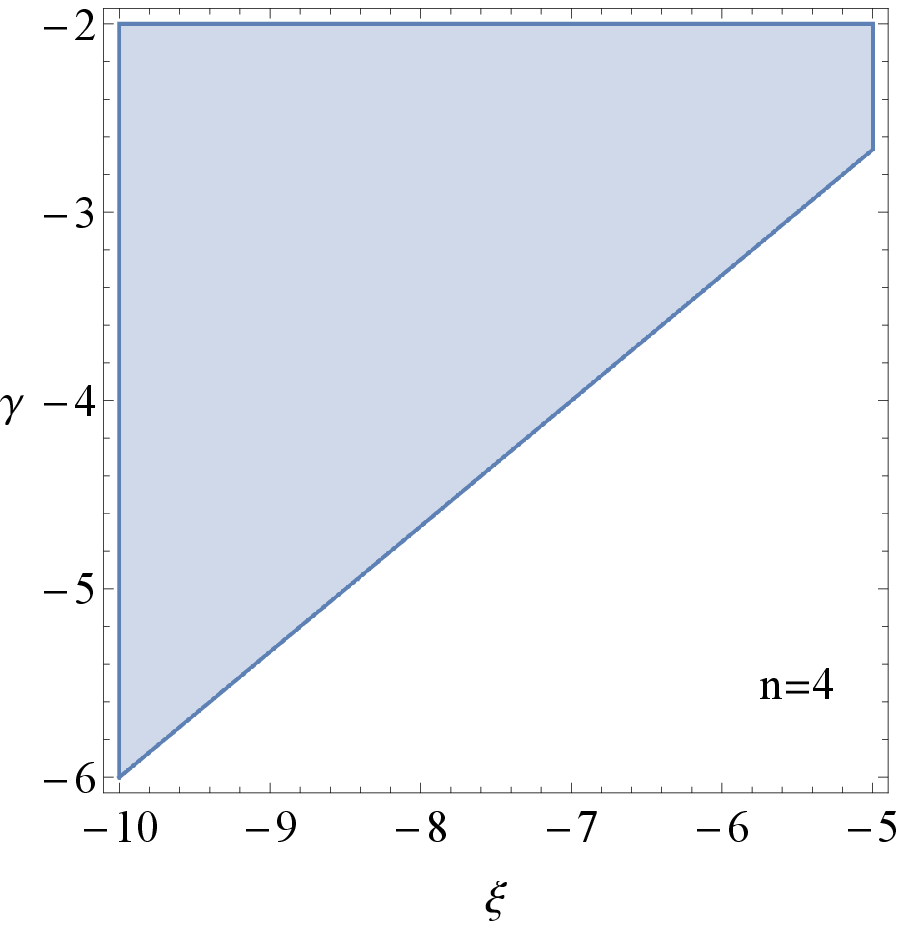}
                \caption{\label{Fig1} Parameter space plot $\gamma$ against $\xi$ showing the stable regions of homogeneous and inhomogeneous scalar perturbations with $n=0,2,3,4$. $n=0$ corresponds to the regions of homogeneous perturbations. These figures are plotted for $\omega=1$, $V_{0}=1$ and $\lambda_{0}=-7$.}
        \end{figure}

\begin{figure}[!htb]
                \centering
                \includegraphics[width=0.45\textwidth]{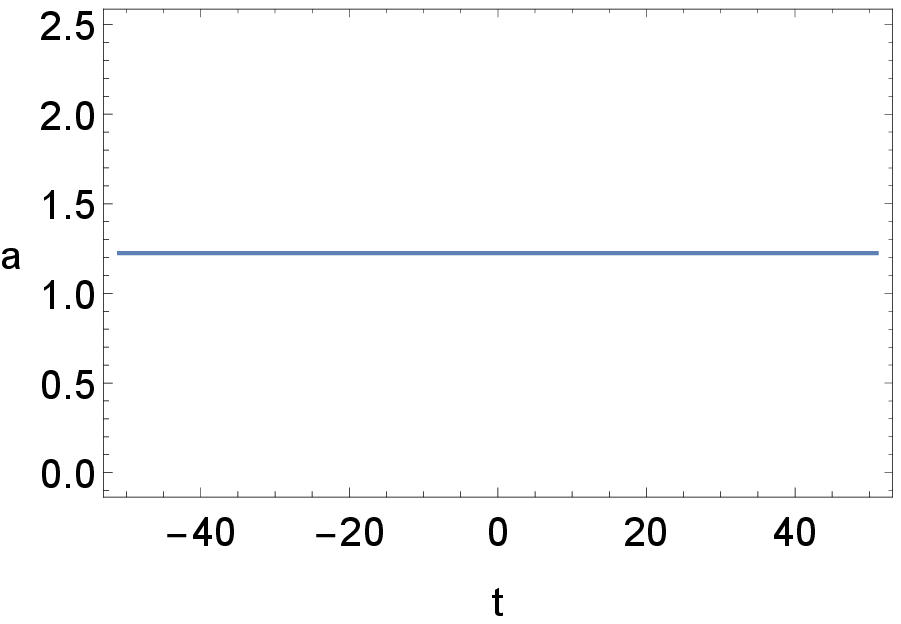}
                \includegraphics[width=0.45\textwidth]{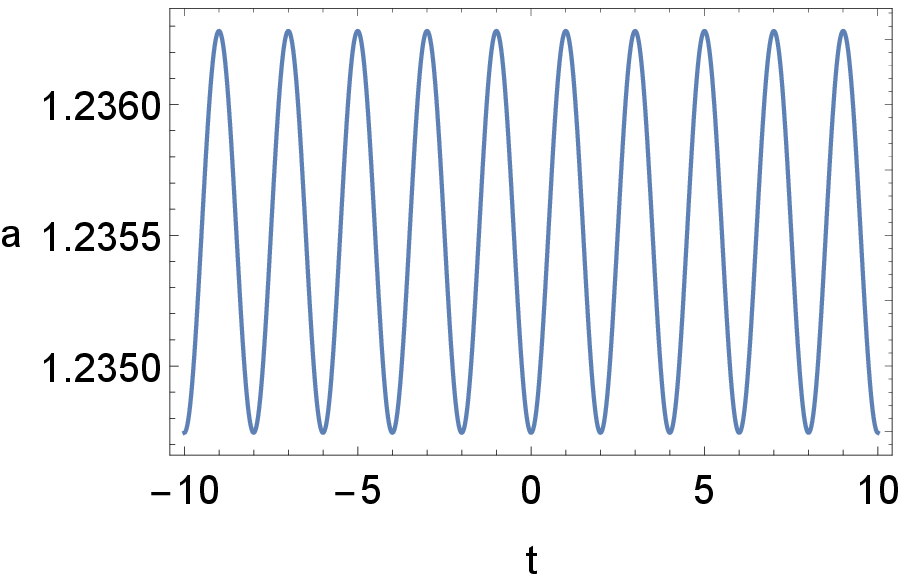}
                \caption{\label{Fig2} Evolutionary curve of the scale factor $a$ with time $t$. These figures are plotted for $\omega=1$, $V_{0}=1$, $\lambda_{0}=-7$, $\xi=-10$ and $\gamma=-5$. The initial value of scale factor $a$ in the left panel is ES solution $a_0$, while a slight deviation of $a_0$ is considered in the right panel.}
        \end{figure}

(iii)Using the conditions Eq.~(\ref{h3}) and ~(\ref{MN}), we can get the stable regions for the ES solutions
\bea\label{ss3}
&&V_{0}>0, \quad \omega>\frac{3}{7}, \quad \xi<-1, \quad \frac{2}{3}(1+\xi)<\gamma<\frac{(1+\xi)(1+3\omega)}{8\omega},\nonumber\\
&&\lambda_{0}<\frac{(4\gamma -3\xi)(1+\omega)V_{0}}{8\omega\gamma-(1+\xi)(1+3\omega)}.
\eea
An example plot is shown in Fig.~(\ref{Fig1}). In this Figure, $n$ is taken to be $n=0,2,3,4$, where $n=0$ represents the case of homogeneous scalar perturbations. Since the stable regions for $n\geq 4$ overlapped with the case of homogeneous perturbations $n=0$, it is found that a minimum overlapped region for all values of $k^{2}\geq 8$ is the region for $n=2$. Thus, there exists a region in which the ES solutions are stable under homogeneous perturbations and all inhomogeneous perturbations. The evolutionary curves of the scale factor $a$ with time $t$ are plotted in Fig.~(\ref{Fig2}). In the left panel, the initial value of the scale factor is ES solution $a_0$, this panel shows that the evolution curve of $a$ parallels to t-axis. In the right panel, we consider the initial value of $a$ deviates slightly from ES solution $a_0$, the result shows that the scale factor oscillates around the ES state. These figures show that the ES solutions are stable in the stable regions.

\section{Leaving the Einstein static state}

In previous sections, we analyze the stability of the ES solutions and find the ES solutions are stable under some certain conditions. Thus, the universe can stay at the stable ES state past eternally. However, a successful emergent scenario not only requires the universe can stay at the stable ES past eternally but also can exit from the static state and evolve to a subsequent inflationary era~\cite{Ellis2004,Ellis2004a}. In this section, we will show its viability.

\begin{figure}[!htb]
                \centering
                \includegraphics[width=0.45\textwidth]{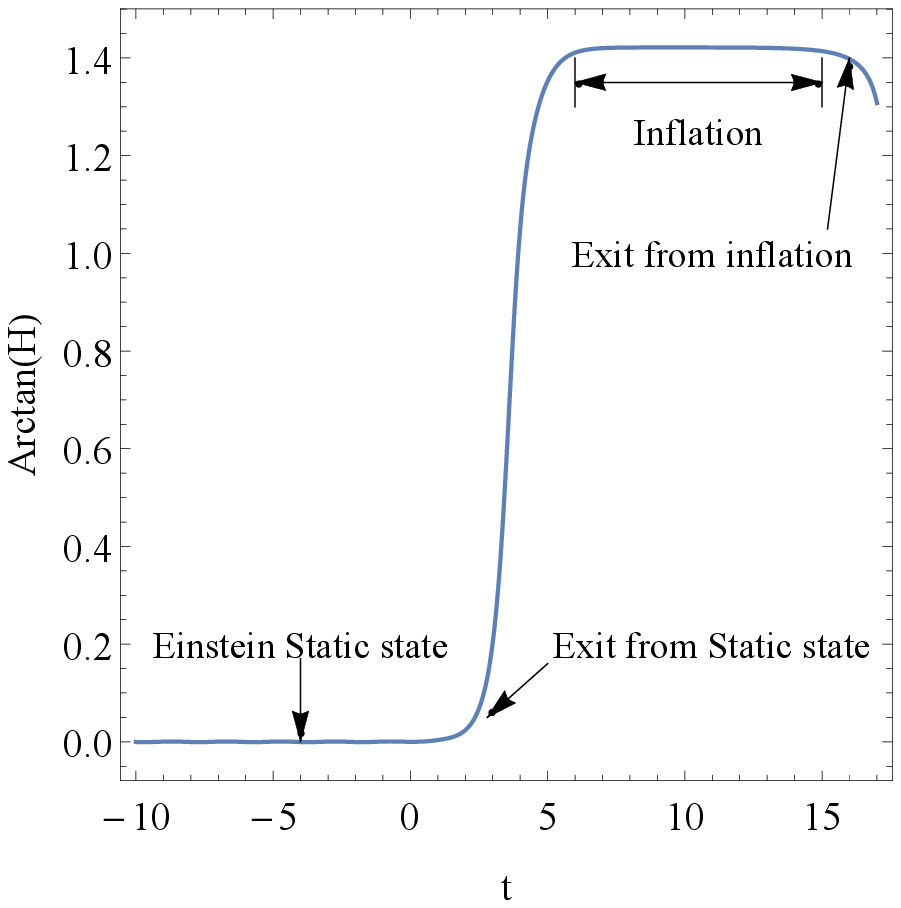}
                \includegraphics[width=0.45\textwidth]{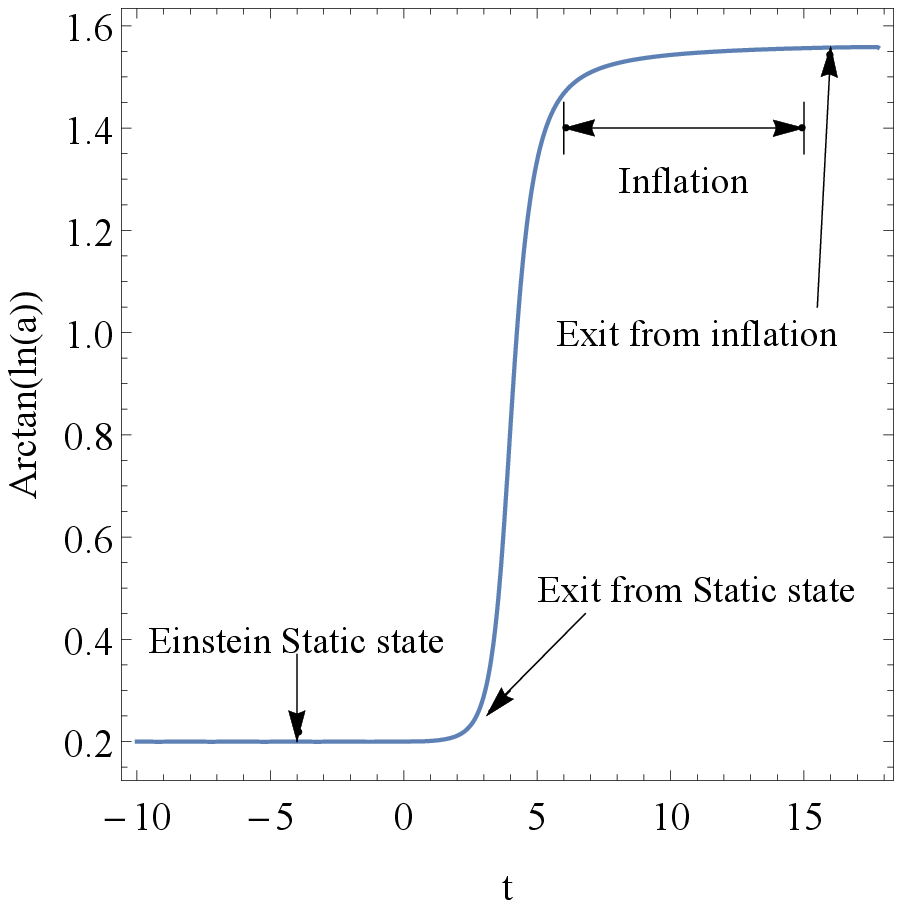}
                \caption{\label{Fig3} Early evolution of the universe. The left panel depicts the evolutionary curve for $H$, while the right one is for $a$.  After the universe leaves from initial static state, it evolves into an inflationary era and then exits from this era. These figures are plotted for $\omega=1$, $V_{0}=1$, $\alpha=12$, $\beta=100$, $\sigma=5$, $\lambda_{0}=-7$, $\xi_{0}=-10$ and $\gamma=-5$.}
        \end{figure}

To achieve this purpose, we assume the potential has the form $V=V_{0}+e^{\phi-\alpha}+\beta[\tanh(\sigma-\phi)-1]$, where $V_{0}$ represents the static value in previous section, $\phi=t$ since $\dot{\phi}^{2}=1$, and $V=V_{0}$ when $t\rightarrow-\infty$. Then as time passes, $V$ departs from $V_{0}$ and the Einstein static condition is broken, the exit from Einstein static universe can be realized. Then, we solve the dynamical equations ~(\ref{00}), ~(\ref{ij}) and ~(\ref{s}) numerically, and the results are shown in Fig.~(\ref{Fig3}) in which we have plotted the early evolution of universe. The left panel shows the evolution of $H$, after it departs from initial value, it evolves rapidly as a constant which represents inflation, and then decreases and exits from inflation. The right panel shows that after the universe leaving from the initial static state, the scale factor $a$ can deviate rapidly from the static state, evolve into a subsequent inflationary era and then exit from inflation.

In standard slow-roll inflation, the e-folding number is used to determine the amount of inflation and is defined as
\beq
N=\ln\Big(\frac{a_{e}}{a_{i}}\Big),
\eeq
where $a_e$ is the scale factor at the end of inflation and $a_i$ represents the one at the beginning of inflation. According to numerical calculation, we estimate the e-folding number as $N \simeq 59$.

In addition, the matter with $\omega=1$ is stiff matter, so an unknown mechanism after inflation should be introduced in which this matter should decay in the usual matter and radiation.

\section{Conclusion}

The mimetic gravity, which is an extension of general relativity, can serve as a model of dark matter and dark energy. It has also been suggested that it allows a bouncing universe to avoid the big bang singularity. In this paper, we have studied the possibility of resolving the singularity problem by using the emergent mechanism in mimetic gravity. We analyzed the stability of ES solutions against scalar perturbations and found that the stable ES solutions exist with the allowed regions of parameters being given in Eq.~(\ref{ss3}). However, since $\xi<-1$ is needed, the ES solutions are unstable for the model with higher derivative terms of the mimetic field~\cite{Chamseddine2014}. And $\gamma<0$ shows the stable ES solutions can not exist in the original mimetic gravity~\cite{Golovnev2014,Chamseddine2014}. In addition, the ES solutions are also unstable for radiation or dust matter as $\omega>\frac{3}{7}$ is required. Then, by assuming the potential $V$ has the form $V=V_{0}+e^{\phi-\alpha}+\beta[\tanh(\sigma-\phi)-1]$, we find that the universe can naturally exit from the initially stable static state, evolve into an inflationary era and then exit from inflation. Thus, in mimetic gravity, the emergent mechanism can be used to resolve the big bang singularity under certain conditions.

\begin{acknowledgments}

This work was supported by the National Natural Science Foundation of China under Grants Nos. 11865018, 11947030, 11865019, the Foundation of the Guizhou Provincial Education Department of China under Grants Nos. KY[2018]312, KY[2017]247, the Doctoral Foundation of Zunyi Normal University of China under Grants No. BS[2017]07.

\end{acknowledgments}

\end{document}